\documentclass[hyperref]{article}
\usepackage{makeidx}
\usepackage[colorlinks=true,citecolor=black]{hyperref}
\usepackage{hyperref}
\usepackage{multirow}
\usepackage{multicol}
\usepackage[dvipsnames,svgnames,table]{xcolor}
\usepackage{graphicx}
\usepackage{epstopdf}
\usepackage{ulem}
\usepackage{hyperref}
\usepackage{amsmath}
\usepackage{amssymb}
\usepackage{enumerate}
\usepackage[a4paper]{geometry}

\usepackage{caption}

\makeatother 

\begin{document}
\captionsetup[figure]{labelfont={bf},name={Fig.},labelsep=period}
\title{\bf\Large How to find a GSMem malicious activity via an AI approach}

\date{}
\author{\large ZHU Wei-Jun, BAN Shao-Huan and FAN Yong-Wen\\
{\sffamily\small School of Information Engineering, Zhengzhou University, Zhengzhou,
450001, China}
}
\maketitle
{\noindent\small{\bf Abstract:}
     This paper investigates the following problem: how to find a GSMem malicious activity effectively. To this end, this paper puts forward a new method based on Artificial Intelligence (AI). At first, we use a large quantity of data in terms of frequencies and amplitudes of some electromagnetic waves to train our models. And then, we input a given frequency and amplitude into the obtained models, predicting that whether a GSMem malicious activity occurs or not. The simulated experiments show that the new method is potential to detect a GSMem one, with low False Positive Rates (FPR) and low False Negative Rates (FNR). }

\vspace{1ex}

\section{Introduction}
Some cyber attacks have occurred during the last three years, which was viewed as impossible in the traditional secure opinion. For examples, an air-gapped computer is no longer secure under some cyber attacks. And there are a number of attack means that allow an attacker to theft information from an air-gapped computer at present. A network weapon called ”suter”, which is developed by the US army, is employed to attack the air-gapped military computers. Using this technique, the Israelis successfully invaded the Syrian air defense radar network and take over its control. As a result, the Israeli Air Force successfully implemented the bombing. 

The successful cases of cyber attacks further stimulate the rapid development of the air-gapped attack technique. The new intrusion techniques are emerging. And this types of attacks has been developed to theft data from a air-gapped computer using the thermal emission \cite{1}, electromagnetic radiation \cite{2}, ultrasonic \cite{3} \cite{4}, or USB devices \cite{5}, respectively. GSMem attack is one of the representative air-gapped attacks. However, the existing Intrusion Detection (ID) techniques can hardly detect GSMem attacks. To address this problem, we introduce a Machine Learning (ML) method to detect GSMem attacks. This is the main contribution of this paper.
\section{Preliminary}
\subsection{The GSMem Attack Model \cite{2}}
Air-Gapped computers are physically and logically isolated from any network. This section describes the principle of ”GSMem” attacks that can steal data from air-gapped computers. ”GSMem” is a distributed malware, which consists of the two parts. One is installed on an air-gapped computer, and it transmits the signal. And the other runs in a mobile baseband processor, so that it can receive the signal. The former is described as Launcher Module (LM), and the latter is called Receiver Module (RM). 
\begin{figure}
	\includegraphics[width= \textwidth]{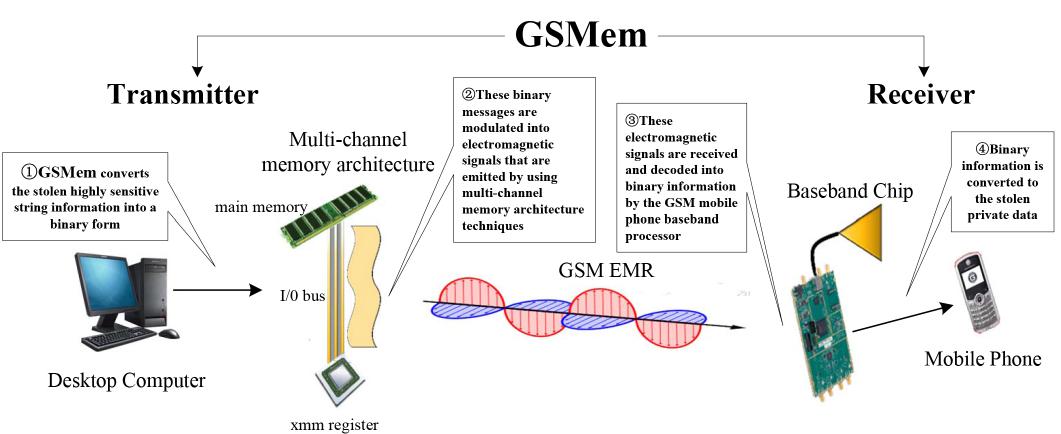}
	\caption{schematic diagram for a GSMem attack}
	\label{fig1}
\end{figure}

Fig. 1 illustrates the core principle of a GSMem attack. The key steps can be depicted as follows.
\begin{enumerate}[1)]
\item The signal modulation program in LM converts the stolen information into a binary string.
\item The signal modulation program in LM periodically calls the Move Double Quadword Non-Temporal (MOVNTDQ) instruction, which is one member of the CPU instruction sets of Single Instruction Multiple Data (SIMD). These calls generate the electromagnetic waves in the GSM band which is compatible with mobile basebands, according to the B-ASK modulation mode. In this way, the binary string is embedded into a sequence of GSM electromagnetic signals.
\item The signal modulation program in LM increases the amplitude of electromagnetic waves using multichannel memory architecture, since multichannel information can transmit the stronger electromagnetic waves. And these amplified electromagnetic signals that carry the binary string will be emitted.
\item RM in a standard GSM mobile phone receives and decodes these emissions of electromagnetic signals, resulting in data containing user privacy.
\end{enumerate}
\subsection{The several ML approaches}
A training set and a test set are needed using ML approaches. In this paper, we will employee the following six ML approaches: Logistic Regression (LR), Random Forest (RF), Support Vector Machine (SVM), Boosted Tree (BT), Back-Propagation Neural Networks (BPNN) and Naive Bayes Classifier (NBC). See Ref.\cite{6} for more details on the various ML approaches. We do not expand it here due to limitation of space.

\section{The ML-based method for detecting GSMem attacks}
\begin{figure}
	\includegraphics[width= \textwidth]{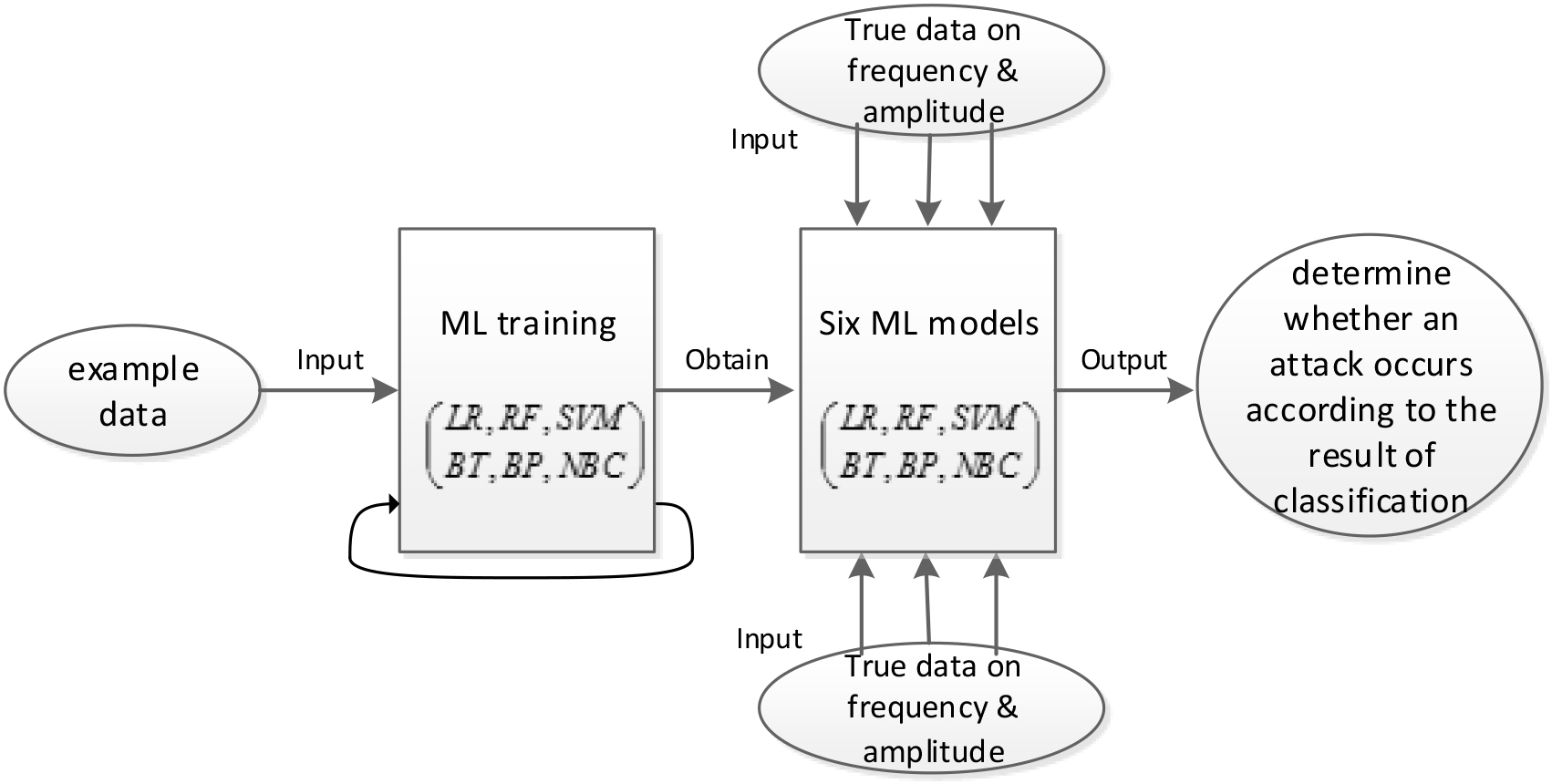}
	\caption{The principle of the ML-based method for detecting GSMem attacks}
	\label{fig1=2}
\end{figure}
Prof. Guri has given a great number of data on the frequency and the amplitude emitted by the electromagnetic waves \cite{2}, which indicate some normal cyber behaviors and the GSMem attacks. With these data at hand, one can train the models that have an ability to predict more GSMem attacks. According to the basic ML principle, we can design a new method for detecting GSMem attacks, as shown in Fig. 2.   

\section{The Simulation Experiments}
\subsection{Experimental Objective}
In section 4, we aim to explore the ability of the new method.
\subsection{Experimental Platform}
The used platform is the following one:
\begin{enumerate}[1)]
\item Hardware: Lenovo, Intel(R) Core(TM) i7-4790, 3.60 GHz, 8 GB RAM. 
\item OS: Windows 7 64-bit. 
\item Tools: GraphLab Create 3.0.1 \cite{7} for running Logistic Regression, Random Forest, SVM and Boosted Tree; JetBrains PyCharm 2016.1.4 \cite{8} for running BP Neural Networks and Naive Bayesian Classifier.
\end{enumerate}
\subsection{Experimental Procedure}
\begin{figure}
	\includegraphics[width= \textwidth]{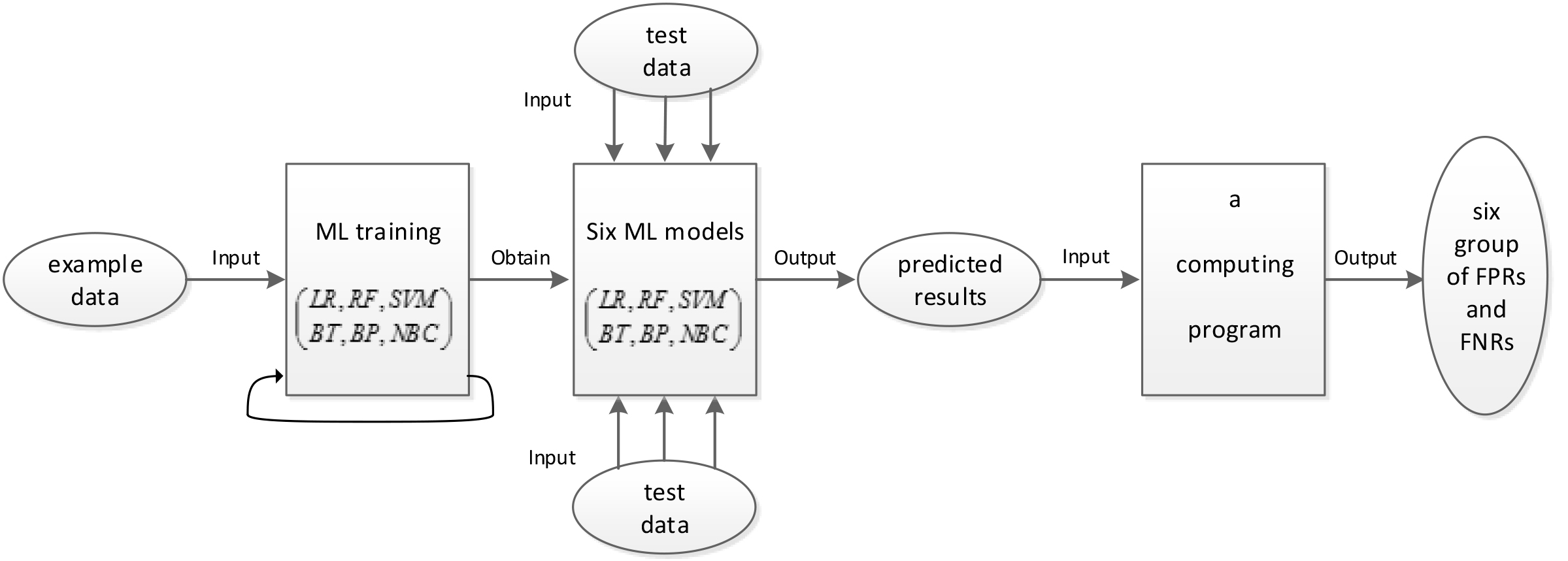}
	\caption{The Principle of the Experiment and Experimental Procedure}
	\label{fig1=3}
\end{figure}
According to the basic principle of the new method, we have employed the above tools to implement LR, RF, SVM, BT, BPNN and NBC, respectively. The key steps are training and prediction. Fig. 3 illustrates the principle of our experiment and experimental procedure.
\begin{figure}
	\includegraphics[width= \textwidth]{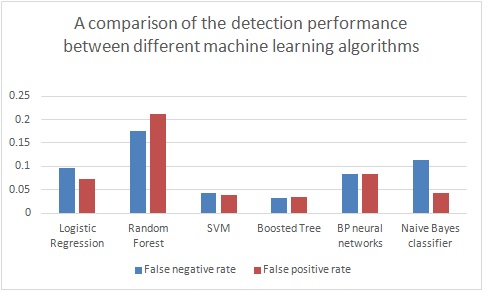}
	\caption{A comparison of FNR \& FPR among different ML-based algorithms}
	\label{fig1=4}
\end{figure}

\subsection{Experimental Results}
Fig. 4 illustrates our experimental results. For all of the six ML-based algorithms, the best FNRs/FPRs are depicted in this figure. 

As shown in Fig. 4, both the SVM-based algorithm and the BT-based one hardly lead to false alarms and leakage alarms. In comparison, it is easier for the RF-based algorithm to lead to false alarms and leakage alarms. This discovery prompts us that the later or the stronger ML approaches do not necessarily bring into the better performance in terms of GSMem detection.

\section{Conclusions}
The main result of this paper is Fig.2, illustrating a ML-based ID method. And the new method is potential to detect actual GSMem attacks by rule and line. This is the benefit of using the new method.
\section*{Acknowledgements}
This work has been supported by the National Natural Science Foundation of China (No.U1204608). We'd like to thank arXiv for giving us a chance to correct some typos in previous version of this paper. For examples, the words "FNR" and "FPR" in Fig.4 are marked incorrectly. And they should be replaced with each other. In addition, the experimental result obtained by a BP neural network is another typo in Fig.4. In this version, we check them carefully and correct them.

\end{document}